\documentclass[12pt]{iopart}
\usepackage{graphicx}
\usepackage{psfrag}

\usepackage{iopams}

\begin{document}

\title[Power Spectra in a ZRP $\ldots$]%
{Power Spectra in a Zero-Range Process on a Ring: Total Occupation 
Number in a Segment}
\author{A G Angel$^{1,2}$ and R K P Zia$^1$}
\address{
$^1$ Physics Department, Virginia Polytechnic Institute and State University,
Blacksburg, VA, 24061, USA}
\address{
$^2$ Department of Computational and Systems Biology, John Innes Centre,
Norwich, NR4 7UH, UK}
\date{\today}

\eads{\mailto{andrew.angel@bbsrc.ac.uk}, \mailto{rkpzia@vt.edu}}
\begin{abstract}
\noindent 
We study the dynamics of density fluctuations in the steady
state of a non-equilibrium system, the Zero-Range Process on a ring lattice.
Measuring the time series of the total number of particles in a \emph{segment%
} of the lattice, we find remarkable structures in the associated power
spectra, namely, two distinct components of damped-oscillations. The
essential origin of both components is shown in a simple pedagogical model.
Using a more sophisticated theory, with an effective drift-diffusion
equation governing the stochastic evolution of the local particle density,
we provide reasonably good fits to the simulation results. The effects of
altering various parameters are explored in detail. Avenues for improving
this theory and deeper understanding of the role of particle interactions
are indicated.
\end{abstract}


\maketitle

\section{Introduction}

For systems in thermal equilibrium, Boltzmann and Gibbs provided a sound
framework which forms part of text-book material nowadays. Specifically, the
probability distribution (for finding the system in any configuration) is
given simply by the Boltzmann factor. In contrast, little is known in
general about systems driven into non-equilibrium steady states, specified
via, say, a set of transition rates that violate detailed balance. To
explore this vast unknown, it is natural to focus on simple solvable models,
with the hope of gaining insight for formulating a general framework for
non-equilibrium statistical mechanics. One paradigmatic model is the
Zero-Range Process (ZRP) \cite{Spitzer70,EH05}, in which particles hop from
one site to the next on a one-dimensional \emph{periodic} lattice (i.e., a
ring of $L$ sites), with a rate that depends solely on $n$, the number of
particles in the originating site.

The ZRP distinguishes itself in at least two ways. Its steady-state
probability distribution not only takes a factorised form, which can be
expressed succinctly in terms of the hopping rate $u\left( n\right) $ (often
easing the computation of various averages). It also exhibits condensation
transitions, even in low dimensions which would not be expected in an
equilibrium system without long-range interactions. In addition to its
utility in the fundamental study of non-equilibrium processes, where it has
been employed to develop a general criterion for phase separation in
one-dimensional driven systems \cite{KLMST02} among other things, the ZRP
has also found success as a minimal model for various real systems including
vehicular traffic \cite{KMH05}, compartmentalised granular gases 
\cite{Torok04} and gel electrophoresis \cite{LK92}.

Though much is known about the ZRP, there is a simple and natural, but so
far unknown, question we may ask. While the total number of particles on the
(finite, periodic) lattice is fixed as the system evolves, the number in a
subsection of, say, $\ell $ sites -- denoted by $N_\ell \left( t\right) $
here -- is a quantity that fluctuates in time $t$. In the steady state, its
time average $\left\langle N_\ell \left( t\right) \right\rangle $ is of
course a constant. Nevertheless, its average power spectrum, $I\left( \omega
\right) \equiv $ $\left\langle \left| \int e^{-i\omega t}N_\ell \left(
t\right) \right| ^2\right\rangle $, is non-trivial and provides information
on the autocorrelation $\left\langle N_\ell \left( t\right) N_\ell \left(
t^{\prime }\right) \right\rangle $. A recent study \cite{AZS07} reported the
presence of interesting oscillations in the power spectra for another simple
paradigmatic non-equilibrium model, the open Totally Asymmetric Simple
Exclusion Process (TASEP) \cite{Spitzer70,Schutz01,Derrida98}. In this work,
we carry out an extensive investigation of the ZRP, with a variety of rates
and a range of ring and subsection sizes ($L$ and $\ell $). In addition to
the oscillations discovered in TASEP on an open lattice \cite{AZS07}, we
found two, ``complementary'' sets of oscillations, one controlled by $L$ and
the other, by $\ell $. If we keep the subsection length fixed and let the
ring size go to infinity, we would have the equivalent of an \emph{open}
lattice (with appropriate input and outflux rates to ensure a finite and
non-vanishing $\left\langle N_\ell \left( t\right) \right\rangle $). In this
sense, our considerations for a closed, periodic lattice can be regarded as
inclusive of open lattices as well.

The dynamics of density fluctuations in the ZRP has been investigated
recently by Gupta, et.\ al.\ \cite{GBM07}, who focused on the variance of the
integrated current through a single site as a function of time. This
quantity also displays damped oscillations, but in the \emph{time} domain.
The period is proportional to $L/v$, with $v$ being the velocity associated
with a fluctuation. Being in the frequency domain, our study complements the
earlier work. One of the components of our $I\left( \omega \right) $
oscillates with period $v/L$, and undoubtedly can be traced to the same
origin. However, in \cite{GBM07}, the damping of the oscillations is not the
main focus and so, diffusive/dispersive properties in the system were not
considered. By contrast, we will analyse such behaviour and find that they
pose the most challenge for theoretical understanding. Further, by
considering a quantity associated with multiple sites ($\ell >1$), we hope
to extract more information about the dynamics of fluctuations.

The presence of $v/L$ may lead some to dismiss these phenomena as ``nothing
but finite size effects.'' However, to understand these effects is very
important, since many such models of non-equilibrium transport are believed
to applicable to physical systems of relatively small $L$s. Thus, in the
examples given above, $L\lesssim 10,000$ for vehicular traffic and gel
electrophoresis. The other popular model of non-equilibrium transport,
TASEP, was first introduced as a possible model for protein synthesis 
\cite{MGP68}, 
where $L$ rarely exceeds $1000$. By contrast, in traditional
macroscopic systems (e.g., in solid state physics), we generally have in
mind $L^3\sim \left( 10^8\right) ^3$. Consequently, the phenomena discussed
here are not merely of theoretical interest, but should be physically
observable.

The paper is organised as follows. Details of the model and its dynamics
will be the focus of the next section. Results of simulations, theoretical
analysis and comparisons are the themes of the following three sections. We
end with a summary and outlook for future research.

\section{Model specification and simulation details}

\begin{figure}[tbp]
\begin{center}
\psfrag{S}{$\ell$} \includegraphics[width=0.3\textwidth]{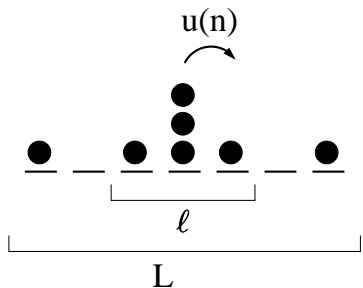}
\end{center}
\caption{General diagram of the system. Particles on a lattice hop to the
rightmost adjacent site with a rate that depends on the number of particles
at the departure site, $u(n)$. The lattice has $L$ sites and periodic
boundary conditions. The quantity of interest is the total number of
particles in a segment of $\ell$ sites of the lattice. }
\label{sysdiagram}
\end{figure}

The ZRP studied here comprises a one-dimensional lattice of $L$ sites with
periodic boundary conditions. A total number of $N$ particles is placed on
the lattice, with no restrictions on the occupation of each site, so that a
configuration of the system is completely specified by the set $\left\{
n\left( x\right) \right\} $, i.e., $n\left( x\right) $ particles being on
site $x$ (with $x=0,...,L-1$). This system evolves through particles hopping
from site to site. Particles hop to the rightmost adjacent site with a rate
that depends only on $n$: $u(n)$. That this hopping rate is \emph{independent%
} of the occupation in any other site in the system gives rise to the ZR
part of ZRP. A basic diagram of the system is shown in Figure~\ref
{sysdiagram}.

Ours is perhaps the simplest ZRP, with the particle landing only in the
nearest neighbouring site. Many more complex cases exist, e.g., having more
than one particle move, landing them in various sites, and inhomogeneous
hopping rates: $u(n, x )$. For a recent review see \cite{EH05}.

Obviously, $\sum_xn\left( x\right) =N$ is a constant in time. The most naive
expectation of our simple ZRP is that, after the system settled into a
(non-equilibrium) stationary state, the average occupation is homogeneous: $%
\left\langle n\left( x\right) \right\rangle =\bar{\rho}\equiv N/L$. Part of
the general interest in the ZRP arises in the existence of a phase
transition, when $\bar{\rho}$ exceeds a critical value $\rho _c$, from such
a homogeneous state to an inhomogeneous ``condensed'' state. In the latter,
all but one site is occupied by an average of $\rho _c$ particles, with the
excess ($N-\rho _cL$) in just a single site. Reminiscent of Bose-Einstein
condensation, this macroscopic fraction is referred to as the condensate.
Unlike the Bose-Einstein case, translational invariance is spontaneously
broken and the condensate can reside in any site. Indeed, in a finite
system, it does disappear from one site and reappear in another on some
interesting time scale \cite{GL05}.

Of the infinitely many functions we can choose for $u\left( n\right) $, it
is the $n\rightarrow \infty $ asymptotic properties that control the
existence of a transition. A typical rate is $u(n)=1+b/n$, which allows a
condensation transition provided $b>2$ \cite{Evans00}. In this paper, we
consider several rate functions, including this one, a constant hop rate,
and $u(n)\propto n$ (which corresponds to having noninteracting particles in
the lattice). Despite the interest in condensation and phase transitions,
here we will focus on the homogeneous phase of the system, in which many
remarkable features already appear. In future studies, we plan to further
investigate the power spectra of systems with a condensate, as well as
lattices with more interesting topologies \cite{AASZ07}.

The system is studied using a simple Monte Carlo algorithm. In each
Monte-Carlo Step (MCS), we make $L$ attempts to move a particle. In an
attempt, a site is selected at random. Provided it contains $n$ ($>0$)
particles, one is moved to the rightmost adjacent site with probability $%
\gamma \, u(n)$, where $\gamma$ is a normalisation factor, $0 < \gamma \leq
1/\max u(n)$. Of course, a particle leaving site $L-1$ is moved to site $0$.
Simulations were typically run for $8\times 10^7$ MCS.

Starting from random initial conditions, we discard the first $10^7$ MCS for
the system to come to the steady state. Subsequently, we focus on a fixed
segment of length $\ell $ sites and record its total occupation, 
\begin{equation}
N_\ell (t)=\sum_{x=0}^{\ell - 1} n\left( x,t\right) \;, 
\end{equation}
every $10$ MCS. Thus, each of our time series consists of $7\times 10^6$
data points, which we regard as $53$ samples of $131072$ points, i.e., $%
t=0,\ldots ,T-1$ with $T=2^{17}$. We choose a power of 2 so that fast
Fourier transform routines can be exploited to compute 
\begin{equation}
\tilde{N_\ell }(\omega )=\sum_te^{-i\omega t}N_\ell (t)\quad ;\quad \omega
=2\pi m/T. 
\end{equation}
Finally, we average over the 53 samples to arrive at the power spectrum: 
\begin{equation}
I(\omega )=\left\langle \left| \tilde{N_\ell }(\omega )\right|
^2\right\rangle \;.  \label{I}
\end{equation}

A typical system size of $L=10000$ sites was used with varying segment
sizes, but most frequently $\ell =1000$. Mostly, we use the Mersenne Twister 
\cite{MN98} random number generator. To rule out systematic errors from
these sources, we have also used other generators (e.g., drand48, ran2,
/dev/urandom), as well as various data segment sizings and output intervals.

\section{Simulation results}

\begin{figure}[tbp]
\begin{center}
\includegraphics[width=0.75\textwidth]{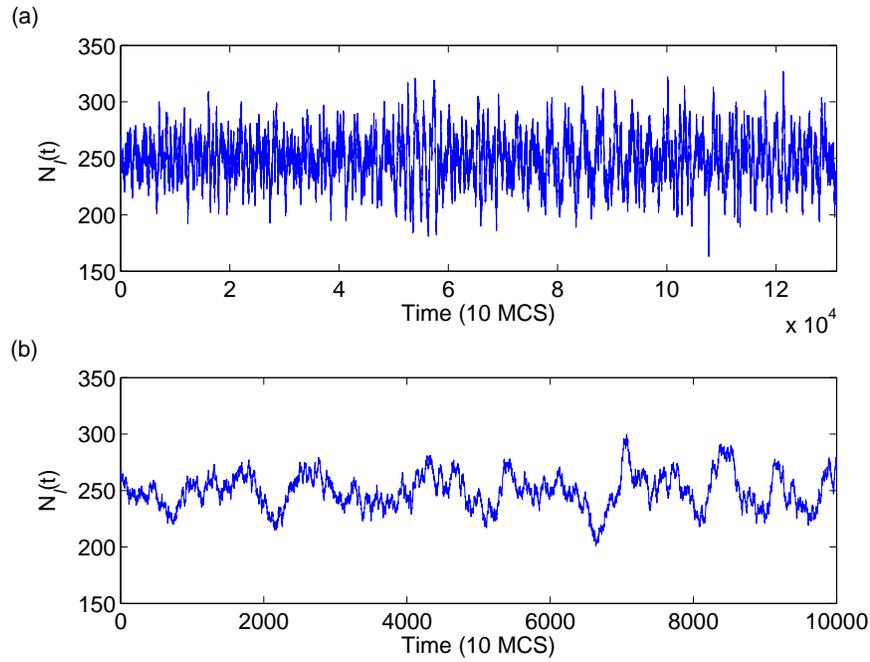}
\end{center}
\caption{A typical trace of the measurement of $N_{\ell}(t)$, the total
number of particles in a section of $1000$ sites in a lattice of $10000$,
from a simulation of a system with $b=4$ at a density of $\rho=0.25$. (a)
One sample set of $131072$ data points. (b) A portion of $10000$ of this
set. }
\label{rawdata}
\end{figure}

\begin{figure}[tbp]
\begin{center}
\includegraphics[width=0.6\textwidth]{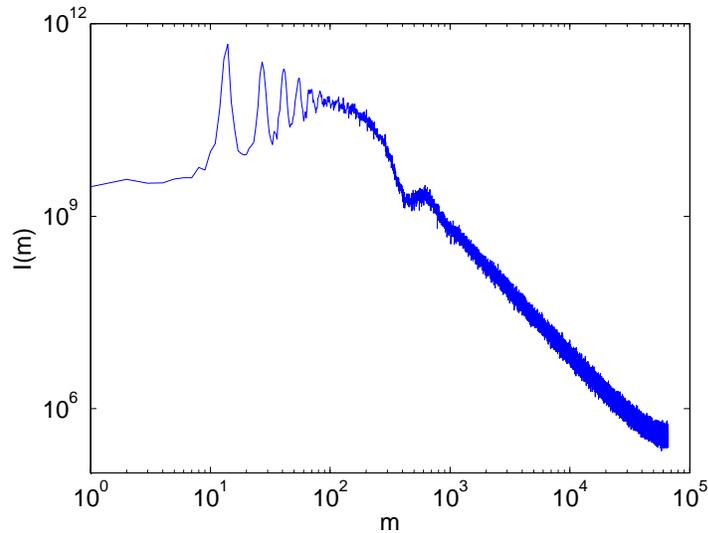}
\end{center}
\caption{A typical result of a power-spectra measurement. Taken from a
system of $L=32000$ sites with a segment of $\ell=1000$, a density of $%
\rho=0.25 $ and hop rate $u(n)= 1+4/n$. The two damped-oscillation
components in the power spectrum can clearly be discerned. The first peak of
the low-$m$ component is at $m=14$ and while the location of the first peak
of the higher-$m$ component is obscured by the oscillation of the other, the
second peak is at $m \approx 650$. }
\label{egspectra}
\end{figure}

A typical trace of $N_\ell (t)$ is presented in Figure~\ref{rawdata},
showing both an entire sample and a magnified view of a section of this
sample. It appears that there is some oscillatory behaviour, but it is
difficult to distinguish from random fluctuations; taking averaged
power-spectra measurements reveals much more in terms of structure. A
typical power-spectra measurement is shown in Figure~\ref{egspectra} for a
system with $L=32000$, $\ell =1000$ and hop rate $u(n)=1+4/n$. It is a
log-log plot and the power-spectrum is plotted against the index, $m$, which
is related to the frequency, $\omega $, through the relation $\omega =2\pi
m/T$. The averaged spectrum displays several prominent features. Two
distinct damped-oscillation components can clearly be seen: one at low $m$
and the other at higher $m$. The former consists of a series of sharp peaks,
with the first peak at $m=14$. Note the \emph{positive} curvature at low $m$%
, a feature notably absent from previously observed power spectra in open
systems \cite{AZS07}. The higher-$m$ oscillations are more subtle, being
obscured partly by the other component. Their character is different,
resembling those observed in \cite{AZS07}, i.e., ``dips'' over a smooth
background. The first ``dip'' can be seen at $m\approx 330$, where the low-$%
m $ oscillations are effectively damped out. For large $m$ the spectrum
tends to $m^{-2}$, characteristic of white noise that might be expected for
frequencies associated with a microscopic time-scale.

The effects of various parameter variations on the structure shown in the
power spectra were studied numerically. These results are discussed in some
detail below.

\begin{figure}[tbp]
\begin{center}
\includegraphics[width=0.6\textwidth]{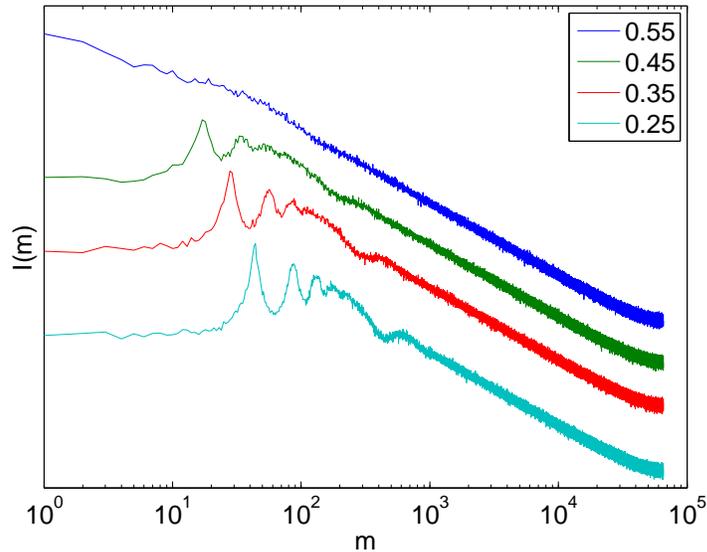}
\end{center}
\caption{Results displaying the effect of varying the density with the
standard hop rate. Data shown are for a system of $L=10000$ sites with a
segment of $\ell=1000$ and hop rate $1+4/n$. The density varies from $%
\rho=0.25$ to $\rho=0.55$ in increments of $0.1$. Note that the data have
been scaled on the $y$-axis for comparison and so the units of this axis are
arbitrary and it is a log-log plot. }
\label{diffdensfig}
\end{figure}

\textbf{Density --- }Examining the changes elicited in the power spectrum
when varying the density in the system it becomes clear that increasing the
overall density lessens and eventually  the 
low-$m$ damped oscillating component, Figure~%
\ref{diffdensfig}. However, the same effect is not conclusively observed for
the higher-$m$ damped oscillations. In fact, simulations at higher values of 
$b$ indicate that this structure remains well into the condensed region, but
at high enough density the oscillations will no longer be apparent. It is
also notable that the removal  
of the low-$m$ oscillations coincides with
the onset of condensation in the system. For the finite systems studied here
condensation does not occur with a sharp transition at the theoretical
critical density (in this case $\rho _c=0.5$) but rather a region of
unstable wandering condensates which become more and more stable as the
density is increased. In the data shown in Figure~\ref{diffdensfig} there is
a moderately-stable wandering condensate for the overall density $\rho =0.55$%
.

\begin{figure}[tbp]
\begin{center}
\includegraphics[width=0.6\textwidth]{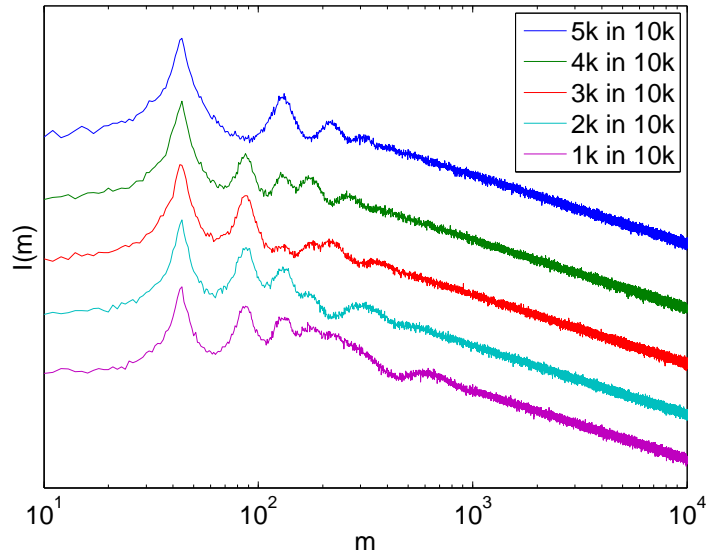}
\end{center}
\caption{Results displaying the effect of varying the segment size with the
standard hop rate. Data shown are for a system with $L=10000$ sites, a
density of $\rho=0.25$ and hop rate $1+4/n$. The segment size varies from $%
1000$ to $5000$ in increments of $1000$. Note that the data have been scaled
for the purposes of comparison, so the units of the $y$-axis are arbitrary and
it is a log-log plot. }
\label{diffsegfig}
\end{figure}

\textbf{Segment Size --- }In Figure~\ref{diffsegfig}, results are shown for
variation in the size of the segment ($\ell $) in the system, keeping the
total number of sites ($L$) and particles ($N$) fixed. Here it can be seen
that increasing the size of the segment causes the higher-$m$ oscillation
component to move to lower and lower $m$ where its interference with the
other component is increasingly apparent. Thus, it is clear that the
structure of the spectrum at high $m$ is due to the size of the segment.
Note that for a segment of $5000$ sites in the ring of $10000$, the higher-$%
m $ component interferes constructively and destructively with alternating
peaks of the low-$m$ component. 

\begin{figure}[tbp]
\begin{center}
\includegraphics[width=0.6\textwidth]{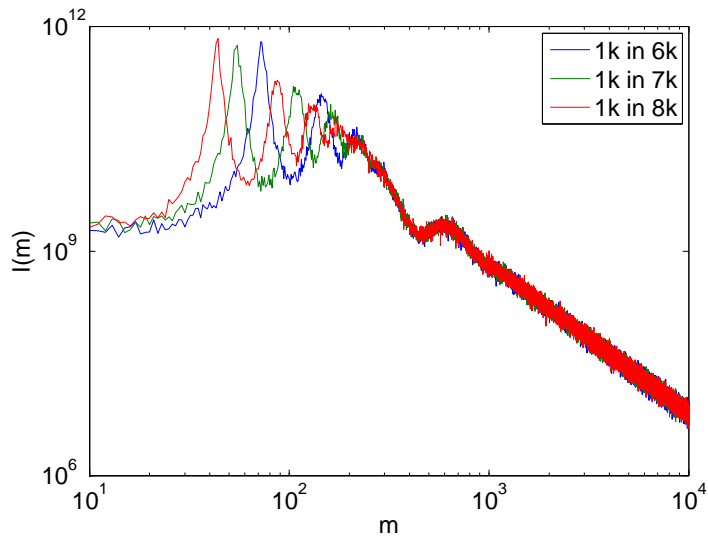}
\end{center}
\caption{Results displaying the effect of varying the lattice size with the
standard hop rate. Data shown are for a system with a segment of $\ell=1000$
sites, a density of $\rho=0.25$ and the hop rate $1+4/n$. The lattice sizes
shown are $6000$, $7000$ and $8000$. }
\label{ringsizefig}
\end{figure}

\textbf{Lattice Size --- }The effect of changing $L$, the lattice size, was
also investigated and results are shown in Figure~\ref{ringsizefig}. It is
clear that at fixed segment size and particle density, increasing the
lattice size changes the low-$m$ damped oscillating component but has little
effect on the higher-$m$ component. In conjunction with the results for
segment size, this leads to the conclusion that the higher-$m$ component is
controlled by the segment size and that the low-$m$ component is controlled
by the size of the lattice.

\begin{figure}[tbp]
\begin{center}
\includegraphics[width=0.6\textwidth]{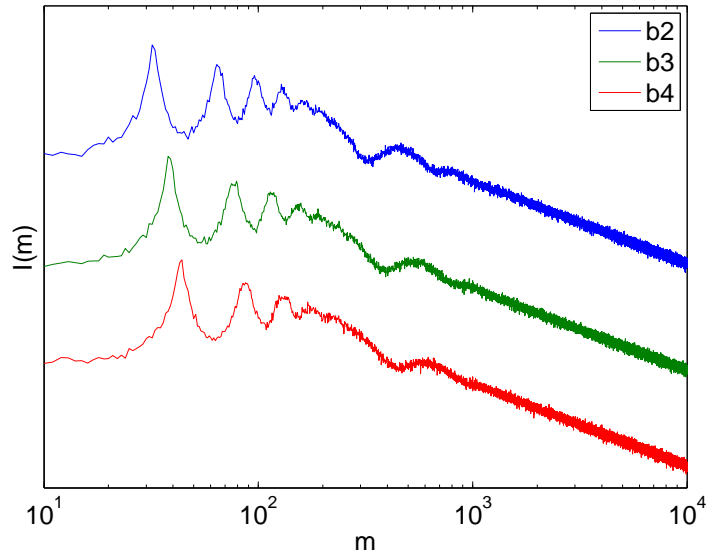}
\end{center}
\caption{Results displaying the effect of varying the parameter $b$ in the
standard hop rate $u(n)=1+b/n$. Data shown are for a system with a lattice
of $L=10000$ sites, a segment size of $\ell=1000$ and density $\rho=0.25$,
with $b$ values $2$, $3$ and $4$. The units of the $y$-axis are arbitrary as
the data have been scaled so that they may be compared on the same graph and
it is a log-log plot.}
\label{diffbfig}
\end{figure}

\textbf{Parameter {\emph{b }}---} Remaining with the standard hop rate, the
effect of varying the parameter $b$ (which can be thought of as controlling
the strength of the interaction) was investigated. Results for this are
shown in Figure~\ref{diffbfig}. It is clear that changing this has an effect
on the location of the peaks in both the low-$m$ and higher-$m$ components.
Also, although not immediately apparent from the data presented, 
it has an effect on the
height of the peaks and the number of clearly resolved peaks.

\begin{figure}[tbp]
\begin{center}
\includegraphics[width=0.6\textwidth]{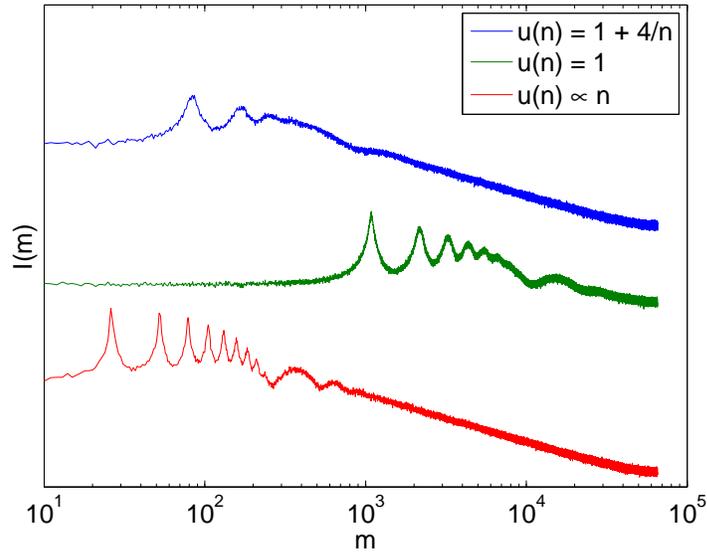}
\end{center}
\caption{Results comparing power spectra from three different hop rates,
from top to bottom: standard $u(n)=1+4/n$; constant $u(n)=1$; and
non-interacting $u(n) \propto n$. The units on the $y$-axis are arbitrary as the
data have been scaled for easy comparison and it is a log-log plot. The data
were taken from a system with $L=1000$ sites, a segment size of $\ell=100$,
a density of $\rho = 0.1$ and the simulations were all run with the same
normalisation, $\gamma$. }
\label{hopratecompfig}
\end{figure}

\textbf{Hop Rate Form} \textbf{--- }Moving on from the standard hop rate, a
comparison of this with constant and non-interacting hop rates is shown in
Figure~\ref{hopratecompfig}. From these results it appears that the two
damped oscillation components are most clearly seen in the case of the
non-interacting hop rate and least clearly seen in the standard hop rate.

\begin{figure}[tbp]
\begin{center}
\includegraphics[width=0.6\textwidth]{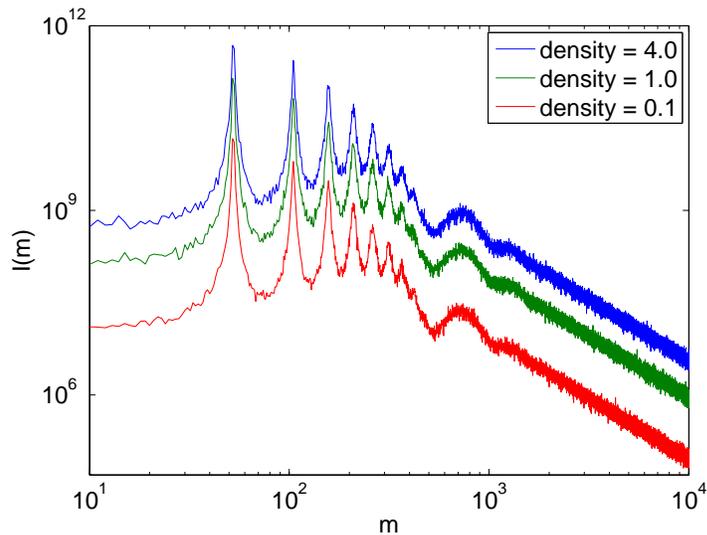}
\end{center}
\caption{ Results displaying the effect of varying the density in the system
with the noninteracting hop rate: $u(n)\propto n$. Data shown are for a
system of size $L=500$ sites with a segment of $\ell =50$ and densities $0.1$%
, $1.0$ and $4.0$. }
\label{nonintdensfig}
\end{figure}

The effect of varying the segment and lattice sizes has much the same effect
on the constant hop rate and noninteracting cases as it did with the
standard hop rate. However, changing the density continues to affect the
power spectra in the case of the constant hop rate, but not for the
noninteracting case. In the latter, changing the density merely changes
magnitude of the power-spectrum, as shown in Figure~\ref{nonintdensfig}.
This suggests that damped oscillations are universal phenomena in
particle-transport systems, though inter-particle interactions will affect
the detailed properties. Indeed, we will show in the next section that a
drift-diffusion type interpretation is quite successful in describing this
phenomenon and that the interaction of the particles controls the values of
the drift and the diffusion.

\section{Theoretical understanding}

Before we present a theory based on the Langevin equation, let us consider a
simple toy model, from which we can gain some insight into the origins of
both types of oscillations. In this toy model, a single particle moves on a
ring of length $L$ with uniform velocity $v_0$, and we seek the power
spectrum of $N_\ell \left( t\right) $, the number of particles in a segment
of length $\ell $. (For simplicity, assume continuous space-time.) Of
course, $N_\ell \left( t\right) $ is trivially a series of step functions,
such that 
\begin{equation}
\partial _tN_\ell \left( t\right) =\sum_{\mu \in \mathbb{Z}}\left[ \delta
\left( t-\mu \tau _L\right) -\delta \left( t-\mu \tau _L-\tau _\ell \right)
\right] 
\end{equation}
where 
\begin{equation}
\tau_{L} \equiv L/v_0 \quad \mbox{and} \quad \tau_{\ell} \equiv \ell /v_0 
\end{equation}
is the time it takes to traverse the ring and the segment, respectively.
Taken over a period of $M$ traversals ($t\in [0,M\tau _L)$, to be
precise), $\tilde{N}_\ell \left( \omega \right) \equiv \int e^{-i\omega
t}N_\ell \left( t\right) $ is given by 
\begin{equation}
i\omega \tilde{N}_\ell \left( \omega \right) 
=
\sum_{\mu
=0}^{M-1}e^{-i\omega \mu \tau _L}\left[ 1-e^{-i\omega \tau _\ell }\right] =%
\frac{1-e^{-i\omega \tau _LM}}{1-e^{-i\omega \tau _L}}\left[ 1-e^{-i\omega
\tau _\ell }\right] \,\,. 
\end{equation}
Thus, 
\begin{equation}
I_{\mathrm{toy}}\left( \omega \right) =\left[ \frac{\sin \left( \omega \tau
_LM/2\right) }{\sin \left( \omega \tau _L/2\right) }\right] ^2\left[ \frac{%
\sin \left( \omega \tau _\ell /2\right) }{\omega /2}\right] ^2 
\end{equation}
For $M$ sufficiently large, the first of these factors gives a series of 
``spikes'' when $\omega$ is an integer multiple of $2 \pi v_0/L$.  Thus, 
these peaks are controlled by the ring size $L$.  By contrast the second 
factor displays an oscillation over a smooth background ($\omega^{-2}$), 
noticeably with zeros at integer multiples of $2 \pi v_0 / \ell$.  
We see that these are 
governed by the segment size $\ell$. 
As a toy model, it also serves as a pedagogical tool. What we
have here is the temporal version of the diffraction pattern from a large ($%
M $) array of slits of width $\ell $, spaced a distance $L$ apart. Of
course, once we add dispersion (velocity here; wavelength in diffraction),
we will see both smoothing of the 
peaks and in-filling of the zeros.

With this insight, we turn to the power spectra here, which can be
reasonably well understood through a Langevin equation for the local
particle density (continuous variable) on discrete space-time: $\rho \left(
x,t\right) $. To connect with the above section, we may think of $\rho (x,t)$
as a kind of coarse-grain average of $n(x,t)$. The starting point is a
discrete continuity equation 
\begin{equation}
\rho \left( x,t+1\right) -\rho \left( x,t\right) =J(x-1,t)-J(x,t)\;,
\label{continuity}
\end{equation}
with $J(x,t)$ being the net local current from site $x$ to $x+1$ at time $t$. 
Clearly, it is controlled by the hop rate $u(x,t)$. Now, as we are considering
power spectra for $\omega >0$, we need only account for the deviations of
this density from the mean, i.e., 
\begin{equation}
\varphi \equiv \rho \left( x,t \right) -
\bar{\rho};\;x=0,1,\ldots,L-1;\;t=0,1,%
\ldots,T-1\;.
\end{equation}
Except for the condensed phase, $\bar{\rho}$ is just the global density of
particles, $N/L$; otherwise, it is $\rho _c$ (apart from the site with the
condensate). The strategy is, for systems far from criticality, these
deviations should be small and their essentials can be understood via an
approximate Langevin equation that is \emph{linear} in $\varphi \left(
x,t\right) $.

Following the standard route, we recognise the deterministic part of $J$ as
a function of $\rho $ and expand that to first order in $\varphi :$ 
\begin{equation}
J_{\det }(x,t)=J_{\det }(\bar{\rho}+\varphi (x,t))=J_{\det }(\bar{\rho}%
)+J_{\det }^{\prime }(\bar{\rho})\varphi (x,t)+...\;.
\end{equation}
Defining 
\begin{equation}
v\equiv J_{\det }^{\prime }(\bar{\rho}) 
\end{equation}
and adding the noisy part of the current, $\eta \left( x,t\right) $, we have 
\begin{equation}
\varphi (x,t+1)-\varphi (x,t)=v\left[ \varphi (x-1,t)-\varphi (x,t)\right]
+\eta (x-1,t)-\eta (x,t)\;.  \label{noisyburgers}
\end{equation}
The noise is assumed to be uncorrelated Gaussians, so that 
\begin{equation}
\left\langle \eta \right\rangle =0;\;\left\langle \eta (x,t)\eta (x^{\prime
},t^{\prime })\right\rangle =A\delta _{x,x^{\prime }}\delta _{t,t^{\prime
}}\,.  \label{noise}
\end{equation}
Here, $A$ is a measure of the strength of the noise, which we regard as a
phenomenological parameter. Before proceeding to the solution, let us
emphasise that this somewhat unusual form of the drift-diffusion equation is
a signature of the ZRP. In many other driven diffusive systems \cite{SZ95},
the current from $x$ to $x+1$ would depend on both $\rho \left( x,t\right) $
and $\rho \left( x+1,t\right) $, so that an additional term involving $%
\varphi (x+1,t)$ will appear on the right hand side of equation (\ref
{noisyburgers}). By contrast, in the ZRP, the jump rates (from $x$ to $x+1$)
depend only on the occupation at $x$.

In this linear approximation, our Langevin equation (\ref{noisyburgers}) can
be easily solved by Fourier methods. Defining 
\begin{equation}
\tilde{\varphi}(k,\omega )=\frac 1{LT}\sum_{x,t}e^{-i(kx+\omega t)}\varphi
(x,t)\,,
\end{equation}
where $k=2\pi j/L$, $\omega =2\pi m/T$ 
($j=0,1,2,\ldots,L-1$ and $m=0,1,2,\ldots,T-1$),
we find the solution easily: 
\begin{equation}
\tilde{\varphi}(k,\omega )=\frac{e^{-ik}-1}{e^{i\omega }-1-v\left[
e^{-ik}-1\right] }\tilde{\eta}(k,\omega )\;.
\end{equation}
Note that, if we keep terms to lowest (relevant) order in $k$ and $\omega $,
the propagator assumes the familiar form, 
\begin{equation}
\frac{-ik}{i\omega +ivk+Dk^2}\,\,, 
\end{equation}
of a drift-diffusion equation with conserved noise: $\partial _t\rho
=D\nabla ^2\rho -v\nabla \rho -\nabla \eta $. For us, the zero-range aspect
of the ZRP imposes a relation between $v$ and the diffusion ``constant'' $D$.

Now, our focus here is the number of particles in a segment and so we
consider 
\begin{equation}
N_\ell (t)=\bar{\rho}\ell +\sum_{x=0}^{\ell -1}\varphi (x,t)\;.
\end{equation}
Carrying out the sum over $x$ of $e^{ikx}$, we find (for $\omega >0$) 
\begin{equation}
\tilde{N}_\ell (\omega )=\sum_k\frac{1-e^{ik\ell }}{1-e^{ik}}\frac{e^{-ik}-1%
}{e^{i\omega }-1-v\left[ e^{-ik}-1\right] }\tilde{\eta}(k,\omega )\;,
\end{equation}
from which we can compute the power spectrum via equation (\ref{I}). Before
comparing such a result to data, we recall that, in the previous study 
\cite{AZS07}, 
the diffusion coefficient seems to be seriously renormalised by
interactions. Anticipating the same behaviour here, we relabel the
coefficient of our $\left[ 1-\cos (k)\right] $ term (i.e., terms even in $k$%
) in the propagator as $D_{\mathrm{eff}}$ --- the effective diffusion
constant --- and regard it as a phenomenological parameter. The resulting
power spectrum, after performing the average over the noise (equation \ref
{noise}), is 
\begin{eqnarray}
\fl I(\omega )\equiv \left\langle \left| \tilde{N}_\ell(\omega) \right| ^2\right\rangle \nonumber
\\
=\frac{2A}{LT}\sum_k\frac{1-\cos (k\ell )}{\left\{ \cos (\omega )-1-D_{%
\mathrm{eff}}(\cos (k)-1)\right\} ^2+\left\{ \sin (\omega )+v\sin
(k)\right\} ^2}  \label{Iofw}
\end{eqnarray}
As we will discuss below, to fit the data well, we must choose values of
both $A$ and $D_{\mathrm{eff}}$ to be far from those naively derived above.
By contrast, we will see that the ``bare'' value of $v$ (i.e., $J_{\det
}^{\prime }(\bar{\rho})$, with $J_{\det }(\rho )$ being the known current
density relationship of the ZRP \cite{Evans00,EH05}) fits the data quite
well. At this stage, there is no good explanation for why $A$ and $D_{%
\mathrm{eff}}$ are significantly ``renormalised,'' while $v$ seems to be
``unscathed.'' Our conjecture is that Galilean invariance imposes a Ward
identity, as in the case of the driven lattice gas \cite{JSLC86}. This is an
avenue which we plan to pursue in the future.

Returning our attention to equation (\ref{Iofw}), we see that it predicts
the locations of the first set of peaks (low $\omega $ ones resulting from 
the sojourn time of a fluctuation around the entire lattice, $L$) of
the power spectra measured from simulation. 
For example, if we consider
small (positive)\emph{\ }$\omega $, we find peaks in $I\left( \omega \right) 
$ whenever $\sin (\omega )+v\sin (k)$ vanishes, i.e., $\omega \simeq 2\pi
jv/L$ ($j=1,2,3,\ldots $ associated with the $k=L-j$ terms). In a similar
way, $\cos (k\ell )$ introduces oscillations on the scale controlled by the
segment length, $\ell $. In particular, the factor $1-\cos (k\ell )$
suppresses the $j^{\mathrm{th}}$ peak if $\ell $ is a unit fraction of $L$,
i.e., $\ell =L/j$. This behaviour, which is reminiscent of interference, is
most pronounced in the case of $j=2$ (top curve $L=10000;\ell =5000$) in
Figure~\ref{diffsegfig}, where the second peak in the other curves is
clearly ``missing.'' A less prominent ``interference'' can be seen for the
third peak in the middle curve ($L=10000;\ell =3000;j\approx 3$). If $\ell
\ll L$, then such effects will be noticeable only at $\omega$'s that are
large compared to those among the peaks. In this regime, the damping is so
severe that the peaks dissolve into a smooth background and the effects of
this factor appear as ``dips.''

Turning to the specifics of the ZRP, the current (in the thermodynamic
limit) is equal to the fugacity $z$ \cite{Evans00,EH05}, so that $v$ will be
given by $\partial z/\partial \rho $. In the case of the standard hop rate, $%
u\left( n\right) =1+b/n$, the $\rho $-$z$ relation is 
\begin{equation}
\rho =\frac z{1+b}\frac{{}_2F_1(2,2;2+b;z)}{{}_2F1(1,1;1+b;z)}\;.
\end{equation}
so that the velocity is 
\begin{eqnarray}
\fl v=(1+b)\,{}_2F_1(1,1;1+b;z)/ \nonumber \\
\left[ {}_2F_1(2,2;2+b;z)+\frac{4z}{2+b}{}_2F_1(3,3;3+b;z)-\frac z{1+b}\frac{%
{}_2F_1^2(2,2;2+b;z)}{{}_2F_1(1,1;1+b;z)}\right] \;.
\end{eqnarray}
Similarly, it is easy to show that, in the other cases, 
\begin{equation}
v=\left\{ 
\begin{tabular}{ll}
$\quad 1$ & for $u\propto n$, i.e., non-interacting particles \\ 
$(1+\rho )^{-2}$ & for $u=\mathrm{const}.$%
\end{tabular}
\right. 
\end{equation}
With these analytic results, we are in a position to examine how well this
theory fits the simulation data.

\section{Comparison between analytic and simulation results}

\begin{figure}[tbp]
\begin{center}
\includegraphics[width=0.6\textwidth]{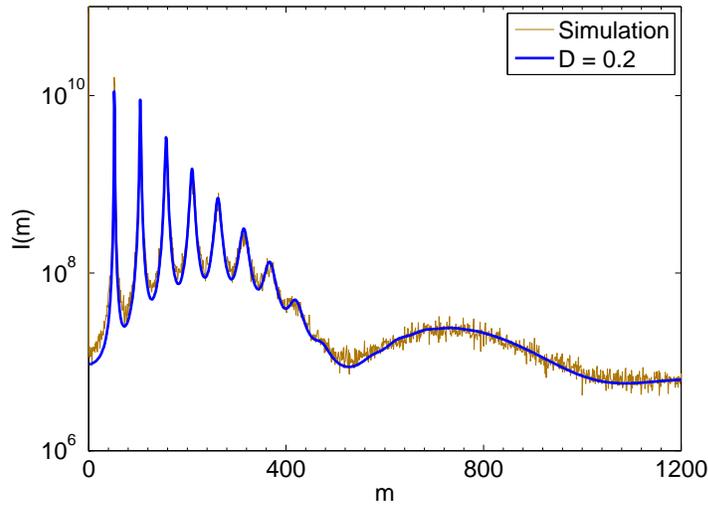}
\end{center}
\caption{Comparison of power spectra taken from simulation data and
generated by the simple theory for the noninteracting system. The system is
a segment of $\ell =500$ sites in a lattice of $L=5000$ with a global
particle density of $\rho =0.1$. Note that in the noninteracting system the
velocity is 1, but the timescale for the simulation (and theory) has been
changed such that its actual value is 0.2. For this system the fit is
apparently good over the entire range and $D_{\mathrm{eff}}=v$.}
\label{nonint_theorysimcomp}
\end{figure}

\begin{figure}[tbp]
\begin{center}
\includegraphics[width=0.6\textwidth]{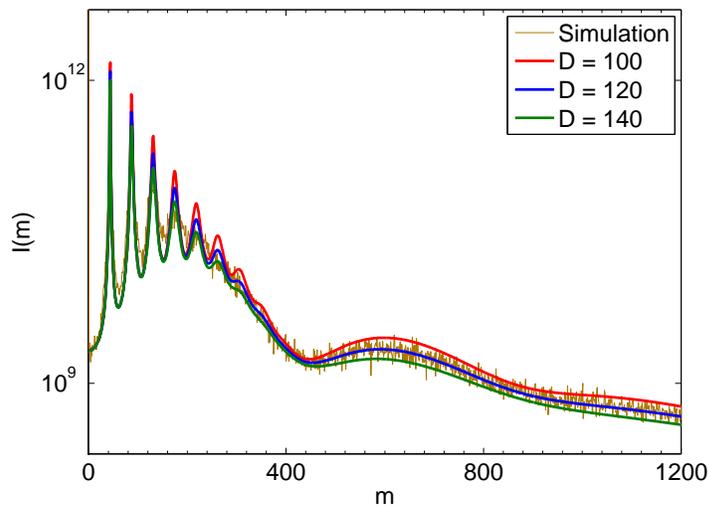}
\end{center}
\caption{Comparison of power spectra taken from simulation data and
generated by the simple theory for the interacting system with the standard
hop rate. The system is a segment of $\ell =1000$ sites in a lattice of $%
L=10000$ with a global particle density of $0.25$ and hop rate $u(n)=1+4/n$.
The system has a velocity of $3.32653$. Here it is apparent that the
effective diffusion constant is a long way removed from the naive expected
value but also that the fit is only really good in the midrange of plot,
around where the second peak due to the size of the segment lies. At the low
end, the diffusion seems to be too small to capture the rapid oscillations
perfectly and at the high end, although not shown here, there is an effect
from a cancellation in the first term of the denominator in (\ref{Iofw}). }
\label{theorysimcompeg}
\end{figure}

There is generally good agreement between the simple theory presented above
and results from simulation, i.e., appropriate choices of the parameters $A$
and $D_{\mathrm{eff}}$ produce reasonable fits. Not surprisingly, the
agreement is best for the noninteracting case. In fact, using $D_{\mathrm{eff%
}}=v$ as predicted by the theory, the match with simulation results over
the entire range of $m$ is quite good --- see Figure~\ref
{nonint_theorysimcomp}. 
For systems
with interacting particles, reasonable fits result only if the effective
diffusion constant, $D_{\mathrm{eff}}$, is drastically changed from the
naively expected value. Moreover, the quality of the the fit is not
uniformly good over all $m$. As an example, in Figure~\ref{theorysimcompeg},
where simulation data (for $L=10^4$, $\ell =10^3$, $u(n)=1+4/n$, and $\rho
=0.25$, below the condensation threshold) are compared with theory plots
with three values of $D_{\mathrm{eff}}$, we see that a value of $120$
matches the data reasonably well. To put this result in context, the naive
(``bare'') value of the diffusion constant is $3.33$, so that we would need
to invoke ``renormalisation effects'' at the level of a factor of $\sim 35$.
Such a sizable ``renormalisation'' is comparable to that observed in the
totally asymmetric simple exclusion process (TASEP) \cite{AZS07}. We believe
the origin is universal -- inter-particle interactions -- and that the
resolution of this issue can be applied to both models. Even treating $D_{%
\mathrm{eff}}$ as a phenomenological parameter, the theory fits well only in
the midrange, with some discrepancies in the low and high $m$ regimes. In
the low-$m$ regime the peaks and troughs are not ideally fitted by the
theory; changing $D_{\mathrm{eff}}$ to match this region typically destroys
the agreement in the mid-range. For $m\gtrsim 1500$ (not shown here), the
lack of agreement is also qualitatively similar, with the addition of 
occasional extra peaks due to a cancellation in the first term in the 
denominator of \ref{Iofw}. 

Since $D_{\mathrm{eff}}$ is seriously modified by interactions, we
investigated systematically the effects of varying various parameters in the
ZRP. In contrast, changing the parameters of the non-interacting system
leaves the $D_{\mathrm{eff}}=v$ relation completely intact.

\begin{figure}[tbp]
\begin{center}
\includegraphics[width=0.6\textwidth]{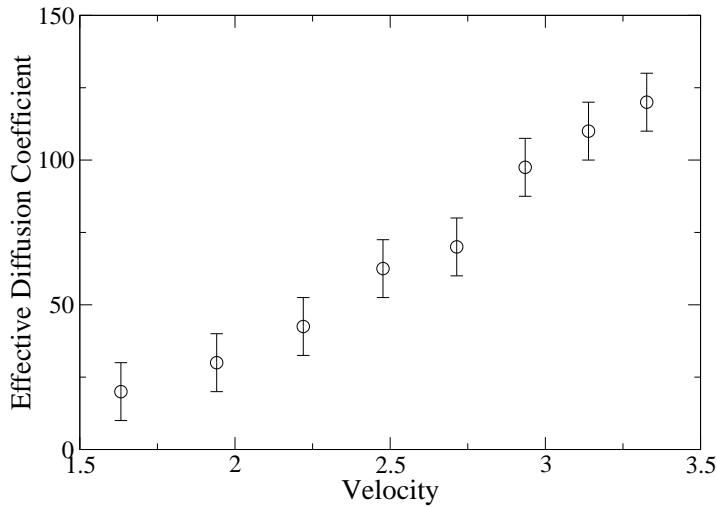}
\end{center}
\caption{Effective diffusion constant plotted against velocity for a varying
value of the standard hop rate parameter, $b$. Data shown are fits between
the simple theory and simulations of a system of $L=10000$ sites with a
segment of $\ell =1000$ at a density of $\rho =0.25$ and $b$ varying from $%
0.5$ to $4.0$ in increments of $0.5$, with low $b$ 
corresponding to low velocity.}
\label{Dv_varyb}
\end{figure}

\textbf{Parameter {\emph{b }}---} The parameter $b$ in the hop rate $u=1+b/n$
can be thought of as a measure of the strength of the interaction:
increasing $b$ first makes a condensation transition possible and then
reduces $\rho _{\mathrm{c}}$, the density at which the transition occurs.
With the density, system size, segment size and normalisation ($\gamma $)
kept constant, the effects of varying $b$ are investigated. Now, neither $v$
nor $D_{\mathrm{eff}}$ varies linearly with $b$, so that we find it more
meaningful to plot $D_{\mathrm{eff}}$ against $v$ (Figure~\ref{Dv_varyb}),
especially to highlight the contrast with $D_{\mathrm{eff}}=v$ for a
non-interacting system. From the figure, $D_{\mathrm{eff}}$ seems to change
with $v$ mostly in a linear way, although the intercept of this linear
component will not go through the origin. Of course, we should expect more
serious non-linear dependence outside the regime shown here.

\begin{figure}[tbp]
\begin{center}
\includegraphics[width=0.6\textwidth]{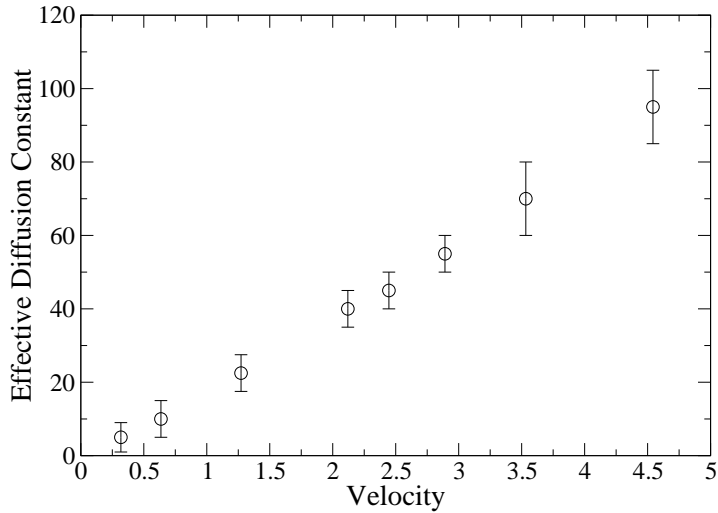}
\end{center}
\caption{ Effective diffusion constant plotted against velocity for a
varying value of the normalisation, $\gamma $. Data shown are fits between
the simple theory and simulations of a system with $L=10000$ sites, a
segment $\ell =1000$ sites, at a density of $\rho =0.1$, hop rate $1+4/n$,
and $\gamma $ values $100^{-1}$, $50^{-1}$, $25^{-1}$, $15^{-1}$, $13^{-1}$, 
$11^{-1}$, $9^{-1}$ and $7^{-1}$; smaller values of $\gamma$ correspond 
to smaller velocities.}
\label{Dv_varynorm}
\end{figure}

\textbf{Normalisation $\gamma $ --- }When simulating systems such as the one
under consideration here, it is common to assign unit probability to the
most likely event, in order to reduce simulation time. This corresponds to
choosing $\gamma =1/\max \left\{ u\left( n\right) \right\} $. For the data
presented in Figure~\ref{Dv_varyb} above, this convention was \emph{not }%
followed. Instead, a single $\gamma $ was used for all the systems (to
ensure a conformity of time, $t$). To explore the effect of varying $\gamma $%
, we also carried out simulations with the usual convention. The behaviour
of $D_{\mathrm{eff}}$ with the velocity was found to change dramatically:
the previously positive trend of $D_{\mathrm{eff}}$ with $v$ was reversed.
To investigate this further, the effect of changing $\gamma $ \emph{without}
other changes was studied. As shown in Figure~\ref{Dv_varynorm}, this gives
an apparently linear relationship between $D_{\mathrm{eff}}$ and $v$. This
is reassuring, since changing $\gamma $ alone should correspond to nothing
more than changing the time scale, while both $D_{\mathrm{eff}}$ and $v$ are
linear in $t$. More puzzling is the value of the gradient in this plot: The
line $D_{\mathrm{eff}}=20v$ fits well inside the error bars. Note that a
factor of $20$ is comparable to the factor $35$ shown earlier. We believe
that, once the origin of the substantial renormalisation of $D_{\mathrm{eff}%
} $ is uncovered, the resolution of this puzzle will follow.

\begin{figure}[tbp]
\begin{center}
\includegraphics[width=0.6\textwidth]{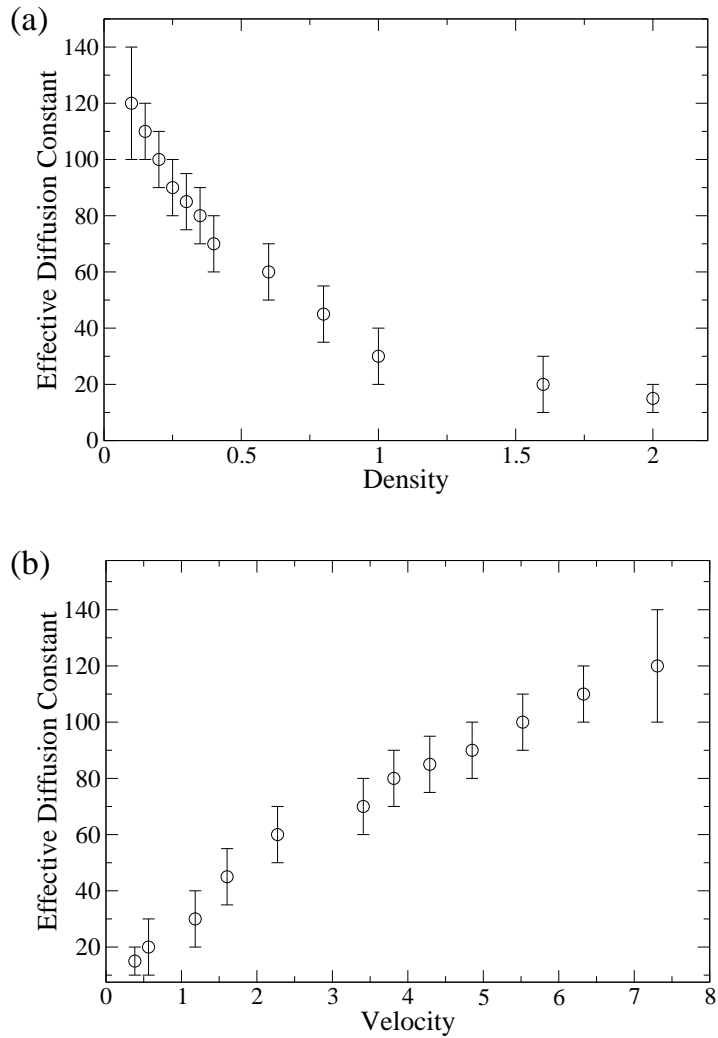}
\end{center}
\caption{(a) Effective diffusion constant plotted against density for a
varying value of the density, $\rho$. A non-linear decrease of the effective
diffusion constant is shown with increasing density. (b) Effective diffusion
constant plotted against velocity for a varying value of the density, $\rho$%
. Densities are $\rho=$ 0.1, 0.15, 0.2, 0.25, 0.3, 0.35, 0.4, 0.6, 0.8, 1.0,
1.6, 2.0, with increasing density corresponding to decreasing velocity. 
For both, data shown are fits between the simple theory and
simulations of a system with $L=10000$ sites, a segment of $\ell=1000$ and a
hop rate of $1+1/n$. }
\label{Drho_Dv_varyrho}
\end{figure}

\textbf{Density --- }With fixed $L,$ $\ell ,$ and $b$, increasing the
overall density ($\rho $) lowers both $v$ and $D_{\mathrm{eff}}$
dramatically. An example of $D_{\mathrm{eff}}$ vs. $\rho $ is shown in
Figure~\ref{Drho_Dv_varyrho} (a). Once we take into account the $v$-$\rho $
relationship and plot $D_{\mathrm{eff}}$ against $v$ (Figure~\ref
{Drho_Dv_varyrho} b), we again recover the line $D_{\mathrm{eff}}\sim 20v$,
with possibly a small negative curvature.

\begin{figure}[tbp]
\begin{center}
\includegraphics[width=0.6\textwidth]{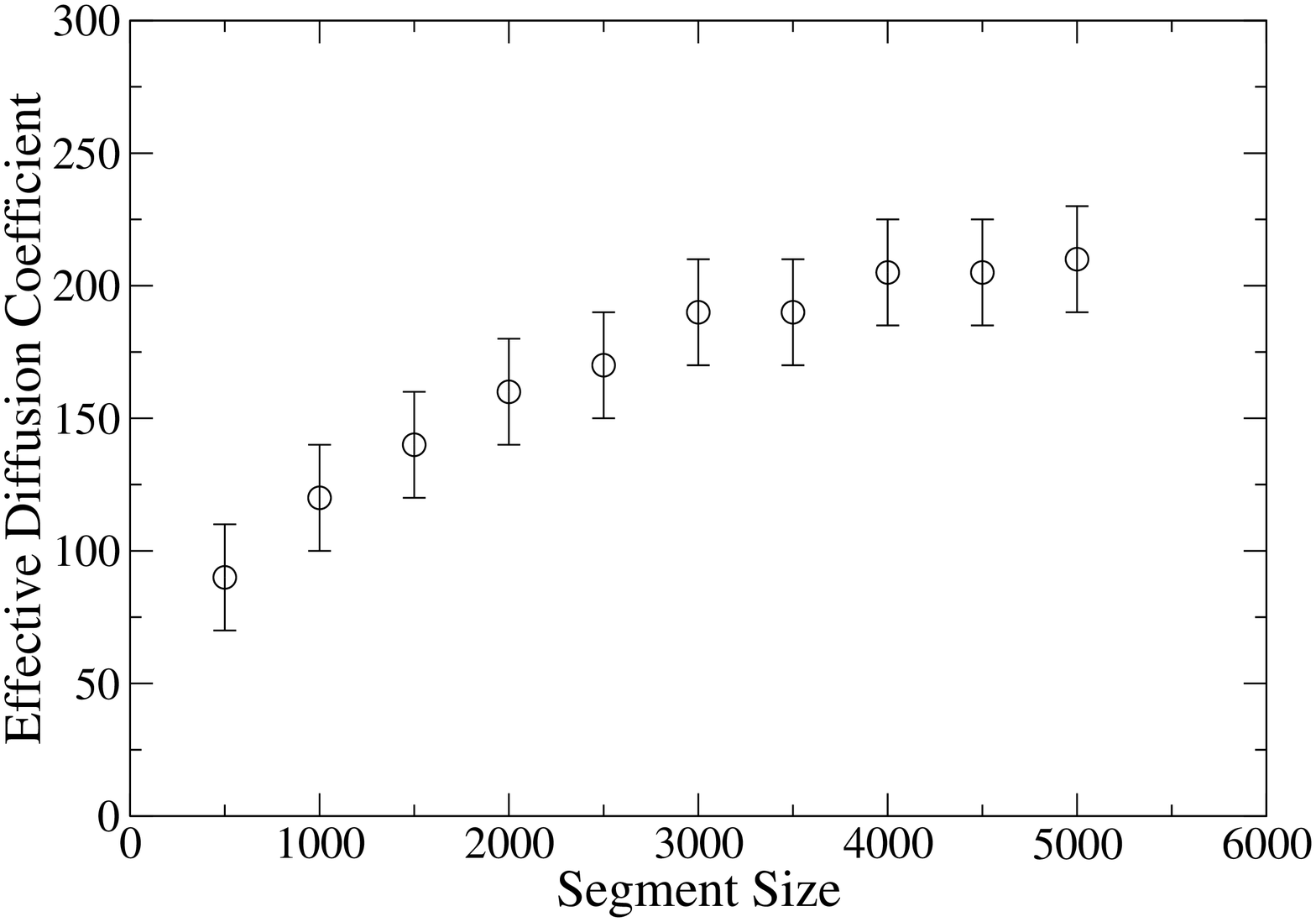}
\end{center}
\caption{Effective diffusion constant plotted against segment size. 
Data shown are fits between the simple theory and
simulations of a system with $L=10000$ sites, a density of $\rho=0.25$ and a
hop rate of $1+4/n$.}
\label{Dseg_varyseg}
\end{figure}

\textbf{Segment Size --- }Changing the segment size ($\ell $) has no effect
on the velocity, but it does have a pronounced effect on the value of $D_{%
\mathrm{eff}}$, as shown in Figure~\ref{Dseg_varyseg}. The relationship is
sub-linear, a behaviour that could be traced to the conserved dynamics. That
is, the particle numbers in a segment ($\ell $) plus those in the
complementary segment ($L-\ell $) is a constant, so that the two averaged
power spectra must be the same. Unless $D_{\mathrm{eff}}\left( \ell \right) $
develops a singularity at the symmetry point, $\ell =L/2$, we must have $%
\partial _\ell D_{\mathrm{eff}}\left( L/2\right) =0$. From this
perspective, a sub-linear variation may have been expected so as to give a
flat profile around $\ell =L/2$. However, it is not obvious why the
relationship should take this form in general.

\textbf{System Size --- }
It is also perhaps surprising that varying the system size does not appear
to change $D_{\mathrm{eff}}$. It is especially so when taken in conjunction
with the fact that changing the segment size does have an effect.

It is best to summarise our findings as follows: Although the observed power
spectra can be reasonably well fitted by the predictions of a linear theory,
we must regard the effective diffusion constant $D_{\mathrm{eff}}$ and the
noise amplitude $A$ as phenomenological parameters. By contrast, the data
is entirely consistent with $v$, the velocity predicted from the theory of
ZRPs. It is clear that the simulation results do not support the prediction
of the ``naive'' theory: $D_{\mathrm{eff}}=v$. In addition, by altering the
control parameters in our study (system size, hopping rate, overall density,
and segment size), $D_{\mathrm{eff}}$ is not only affected dramatically, but
also in such a way that its relationship with $v$ changes. At this stage,
none of these features are well understood.

\section{Conclusion}

In this paper, we studied the dynamics of fluctuations in the
non-equilibrium steady state of the zero-range process (ZRP). Specifically,
we collected time series of the number of particles in a contiguous segment
of a ring lattice, and computed their average power spectra, $I\left( \omega
\right) $. We found interesting structures in $I\left( \omega \right) $,
namely, two distinct damped-oscillation components. The small $\omega $
component consists of narrow peaks over a smooth background. The other
component resembles broad dips, similar to those observed in the power
spectra of open TASEP \cite{AZS07}. The origins of these can be traced to,
respectively, the time it takes a fluctuation to travel around the ring and
the time for traversing the segment.

We presented a simple toy model, to shed some light on these two types of
oscillations: a single particle moving ballistically around a ring in
continuous space-time. The time series of the ``total particle occupation''
in a segment is just a periodic square wave, so that its Fourier transform
is the product of a comb and a sinc function, controlled respectively by the
ring and segment lengths. Diffusion and noise would broaden the peaks of the
comb and fill in the zeros of the sinc function, so that the oscillations
will appear damped. Let us emphasise that these oscillations are controlled
by the system and segment sizes, so that they are absent from the usual
autocorrelation function (for particles at one site, in the thermodynamic
limit). Based on the insight from the toy model, we believe these features
are universal for the power spectra of all \emph{finite} driven diffusive
systems.

At the quantitative level, the observed $I\left( \omega \right) $ can be
fitted quite well by a somewhat more sophisticated approach, based on a
Langevin equation for the local particle density. Focusing on systems far
from criticality, we were motivated to linearise this equation about the
average density. The solution of such an approximate equation, even if we
account for discrete space-time, is easy. However, except for the case with
non-interacting particles, not all the parameters of this simple theory fit
the simulation results. In particular, by identifying the even/odd parity
(i.e., $\nabla \Leftrightarrow \pm \nabla $) terms with diffusion and drift,
we assign the parameters $D_{\mathrm{eff}}$ and $v$, respectively. Good fits
can be achieved only when $D_{\mathrm{eff}}$ is chosen to be considerably
larger than the naively predicted value. By contrast, $v$ from the simple
theory appears to be adequate. At present, we can only present the
dependence of $D_{\mathrm{eff}}$ on various control parameters as
phenomenological results from our extensive simulation studies. In the same
vein, the relationship between $D_{\mathrm{eff}}$ and $v$ (for the range of
systems we considered) was seen to deviate significantly from the naive
theoretical prediction of $D_{\mathrm{eff}}=v$. Such ``anomalies''
associated with $D_{\mathrm{eff}}$ were also observed in similar studies of
another system \cite{AZS07}. Another puzzling aspect here is that $D_{%
\mathrm{eff}}$ appears to depend more strongly on the segment size ($\ell $)
than on the system size ($L$). This feature may indicate that the
introduction of an effective diffusion constant to a linear Langevin theory
is not an entirely satisfactory treatment. While this approach is somewhat
successful at fitting individual power spectra, it leaves much room for
improvement, as we seek better understanding and a comprehensive theory.

Although we have ruled out the possibility that these anomalies are due to a
systematic effect of our random number generator (by using different
generators and choosing parameters to be incommensurate with each other), we
have not considered alternative simulation methods. Two such alternatives
come readily to mind. One is kinetic Monte Carlo \cite{KMCAlgos}, where an
appropriate event is chosen at each update and the time advanced according
to a Poisson distribution. Another alternative would be to try a common
method of reducing simulation times which is to pick a \emph{particle} at
random and hop with an appropriate probability. Fundamentally, we believe
that inter-particle interactions are responsible for large deviations from
the simple linearised theory presented here, rather than some subtle effect
due to the details of the particular dynamics we chose.

It is interesting that similar structures in the power spectra have been
observed in two of the most simple models for non-equilibrium systems,
namely TASEP and ZRP. Oscillations seen in the variance of the integrated
current at a single site in the \emph{time} domain \cite{GBM07} are also
undoubtedly related. These suggest that damped-oscillatory behaviour is
universal for finite systems driven out of equilibrium. Further
investigations to place this notion on a sound foundation would be
worthwhile. It is clear that fluctuations in non-equilibrium steady states
are non-trivial and their dynamics induce interesting behaviour. This study
has shown only a limited view in a small corner of this vast area. Even
within this corner, there is room for improvements, especially in more
complete analytic theory. We hope that a better understanding of this
particular problem will lead to deeper insights into the nature of
fluctuations in physical systems driven far from thermal equilibrium.

\ack

This work was supported in part by the US National Science Foundation
through DMR-0705152. We thank J. J. Dong, V. Elgart, S. Mukherjee, B.
Schmittmann for illuminating discussions.

\end{document}